\documentclass[aps,pra,twocolumn,superscriptaddress]{revtex4-2}
\usepackage{amsfonts}
\usepackage{amsmath}
\usepackage{mathrsfs}
\usepackage{amssymb}
\usepackage{graphicx}
\usepackage{float} 
\usepackage{bm}
\usepackage{color}

\usepackage[colorlinks=true,linkcolor=blue,anchorcolor=blue,citecolor=blue,urlcolor=blue]{hyperref}

\newcommand{\ket}[1] {\left| #1 \right\rangle}

\begin{document}

\title{Tunable nonlinear Landau-Zener tunnelings in a spin-orbit-coupled spinor Bose-Einstein condensate}

\author{Zhiqian Gui}
\affiliation{Institute for Quantum Science and Technology, Department of Physics, Shanghai University, Shanghai 200444, China}

\author{Jin Su}
\email{littlesujin@163.com}
\affiliation{Department of Biomedical Engineering, Changzhi Medical College, Changzhi 046000, China}

\author{Hao Lyu}
\email{hao-lyu@oist.jp}
\affiliation{Quantum Systems Unit, Okinawa Institute of Science and Technology Graduate University, Onna, Okinawa 904-0495, Japan}

\author{Yongping Zhang}
\email{yongping11@t.shu.edu.cn}
\affiliation{Institute for Quantum Science and Technology, Department of Physics, Shanghai University, Shanghai 200444, China}

\begin{abstract}
Nonlinear Landau-Zener tunneling is an important nonlinear phenomenon. We propose to stimulate the nonlinear tunneling in a spin-orbit-coupled spinor Bose-Einstein condensate. The system provides an experimentally tunable nonlinearity as well as multiple avoided crossings with tunable gap size in nonlinear dispersion relations. The nonlinearity generates tilted cusp and loop structures around the avoided crossings, and the physical consequence of these nonlinear structures is the nonlinear Landau-Zener tunneling. 
The spin-momentum locking induced by the spin-orbit coupling leads to the time-resolved observation of the nonlinear tunneling by measuring atom populations. 
%The spin-momentum locking of the spin-orbit-coupled dispersion leads to the time-resolved observation of the nonlinear tunneling by measuring the populations. 

\end{abstract}

\maketitle

\section{Introduction} 
\label{introduction}

Landau-Zener tunneling (LZT), occurring nonadiabatically through an avoided level crossing, is a fundamental quantum phenomenon~\cite{Zener,Zagoskin}.
The transition probability is determined by the celebrated Landau-Zener formula and has an exponential feature related to the competition between the avoided crossing gap and the sweeping rate. 
Due to its fundamental importance, LZT has been experimentally implemented in a variety of systems, such as solid-state devices and photonic systems, and a broad range of applications have been found~\cite{Fuchs,Gaudreau,Higuchi,Autti,WangS,Berngruber}.

Highly tunable ultracold atoms provide an ideal arena to investigate quantum dynamics and many-body physics~\cite{Pitaevskii}. Loading atomic Bose-Einstein condensates (BECs) into optical lattices can generate tunable avoided crossings of Bloch energy bands, and the acceleration of optical lattices can induce atomic LZT through the crossings~\cite{QianNiu1996}. Atomic LZT has been experimentally observed in different accelerating optical lattices~\cite{Morsch,Salger}. An intriguing application of the LZT in experiments is to implement St\"uckelberg interferometry with the help of closely related Bloch oscillations~\cite{Kling2010}. Further experimental progresses make the observation of atomic LZT time-resolved~\cite{Zenesini2009,Tayebirad2010}.

On the other hand, BECs feature nonlinearities induced by atomic interactions, and their effects on LZT become an interesting research direction.  The nonlinearity can make the Landau-Zener transition from the lowest to the first excited Bloch bands to be asymmetric with the reversed transition~\cite{Jona-Lasinio,Konotop2005}. When the nonlinearity dominates over the gap size of the optical-lattice-induced avoided crossing, a loop energy structure adhering to the one band of the crossing appears inside the gap~\cite{Wu2002,Diakonov,Machholm,Machholm2004,Seaman,Koller,Chestnov,YeQ}. 
Loop dispersions have been also used to study hysteresis in a ring BEC~\cite{Eckel}.
Applying an acceleration force which pushes atoms to move along the looped avoided crossing  gives rise to nonlinear Landau-Zener tunneling (NLZT), which was first revealed by Wu and Niu~\cite{Wu2000,Wu2003}. When atoms reach the loop edge, there is no further trajectory for them to go due to the hysteretic characteristic of the loop~\cite{Mueller}.
% The motions along the looped avoided crossing give rise to nonlinear Landau-Zener tunneling (NLZT) which was first revealed by Wu and Niu~\cite{Wu2000,Wu2003}. An acceleration force pushes atoms to move along the looped bands, and when atoms reach the loop edge, there is no further trajectory for them to go due to the hysteretic characteristic of the loop~\cite{Mueller}.
Consequently, atoms must jump to neighboring bands, so that the tunneling can occur. The general theory of NLZT has been provided in Ref.~\cite{LiuJ}, which indicates that the transition probability of NLZT does not follow the Landau-Zener formula and it becomes nonexponential. Another outstanding feature of NLZT is that the transition happens even in the adiabatic limit of the sweeping rate due to the breakdown of adiabaticity~\cite{Wu2000,LiuJ,Witthaut}. The loop-induced NLZT has been experimentally observed~\cite{ChenYA}. Its nonexponential transition behavior has been recently confirmed in experiments using an accelerating optical lattice~\cite{Guan,Hardesty-Shaw}.

NLZT becomes a prototype of nonlinear tunneling physics and attracts great research interests.  
For example, it plays an important role on Josephson tunneling in a double-well potential~\cite{Sakellari},
quantum control of NLZT by a periodic driving field shows interesting properties~\cite{ZhangQ,Luoxiaobing2008}, and NLZT was analyzed in the presence of higher-order dispersion~\cite{Xutianfu2023}. 
NLZT has been generalized to other nonlinear systems, such as nonlinear waveguide arrays~\cite{Khomeriki2010} and  superfluid Fermi gas~\cite{Watanabe}
It has also been proposed that NLZT can be induced in complicate avoided crossings involving three energy levels such as a triple-well trapped BEC~\cite{WangGF}. 
More recently, strong nearest- and next-nearest-neighbor interactions are considered in the triple-well trapped Rydberg-dressed BEC, and the resultant exotic loop structures generate intriguing NLZTs~\cite{WeibinLi,McCormack}.
In addition, nonreciprocal NLZT were revealed in non-Hermitian systems~\cite{WangWY,Cao}, 
and ring-shaped dispersions and corresponding NLZTs were studied in Floquet-engineered systems~\cite{LyuG}.

The experimental implementations of synthetic spin-orbit coupling in ultracold atoms~\cite{Lin,Cheuk,Wang,JiS,Campbell,Khamehchi,Mossman,HuangL} bring ultracold atomic physics into a new era~\cite{Goldman,Zhai2015,Zhang}. The spin-orbit-coupling-engineered single-particle dispersion relation has fundamentally interesting features. %(1) 
It possesses multiple energy minima which provide many possibilities for bosonic atoms to condense~\cite{LiY,Lan,Sun,Yu,Martone,Yani2023}. Therefore, the spin-orbit coupling enriches ground-state phase diagram and makes it possible to generate exotic superfluid phases and elementary excitations~\cite{Martone2012,ChenY}. %(2) 
For the spin-1/2 case, there is a spin-orbit-coupled (SOC) avoided crossing in the dispersion,
which can introduce topology into ultracold atoms~\cite{SOCtopology}, and the existence of gap solitons can also be hosted in the gap~\cite{gapsoliton1,gapsoliton2}. 
In a remarkable experiment~\cite{Olson},  the avoided crossing was used to observe completely tunable  LZT,
since the gap size, initial state preparations and the sweeping rate related to the acceleration force are fully adjustable. 
Due to the spin-momentum locking, the time-dependent measurement of the spin polarization makes the observation of the LZT time-resolved. Later, the LZT through the SOC avoided crossing is used to realize St\"uckelberg interferometry by the same experimental group~\cite{Olson2017}. The experiments~\cite{Olson,Olson2017} and theories~\cite{XiongB,Llorente} show that a SOC BEC with negligible nonlinearity is a perfect platform to stimulate LZT. Considering nonlinearity, Ref.~\cite{Zhang2019} demonstrates that a loop structure emerges inside the SOC avoided crossing only when the difference between inter- and intra-component interactions exceeds the gap size. However, the required interactions are not experimentally preferable in the $^{87}$Rb BECs where the spin-orbit coupling is usually implemented.

In the present paper, we reveal that a SOC spinor BEC offers an experimentally accessible platform for studying NLZT. The SOC spinor BEC with a highly tunable quadratic Zeeman effect has been experimentally realized in $^{87}$Rb atoms~\cite{Campbell}. NLZTs are completely tunable and experimentally realizable in the SOC spin-1 BEC. 

(i) Avoided crossings have rich choices  in the SOC  dispersion. There are multiple SOC avoided crossings in the dispersion of the spin-1 spin-orbit coupling. Two of them are induced by a two-photon transition, and the third one originates from a four-photon transition. The gap sizes of these crossings are tunable, and especially the gap of the third one may be very small, which is benefit for NLZTs. 

(ii) The nonlinearity is tunale.  The emergence of loop-like structures inside these avoided crossings only depends on the spin-spin interaction $c_2$ and is completely irrelevant to the density-density interaction $c_0$.  This unique characteristic ensures experimental accessibility. Physics of a spinor BEC are usually determined by the ratio of the spin-spin to density-density interactions, which is very small unless the Feshbach resonance technique applies~\cite{Stamper-Kurn}. 
For the emergence of the loop-like structures, the decoupling of the spin-spin interactions from the density-density part makes it possible to adjust the spin-spin interaction by changing atomic density. It is crucially important for $^{87}$Rb spinor BECs due to their imperfection of Feshbach resonances. 

(iii) Other tunable parameters and time-resolved measurement can be realized. The experiments~\cite{Olson,Olson2017} on the conventional LZT in SOC BECs have already witnessed tunable parameters and the time-resolved measurement of tunnelings due to the spin-momentum locking in the SOC dispersion.

The remainder of the paper is outlined as follows.
In Sec.~\ref{dispersion}, we show the nonlinear dispersion of a SOC spin-1 BEC. The tilted cusps and loop structures can emerge around the SOC avoided crossings.
These exotic structures can be captured by an effective two-level model. Furthermore, the quadratic Zeeman effect provides an essential tool to adjust these nonlinear structures. 
In Sec.~\ref{NLZ}, we study NLZTs with different well-prepared initial states in the presence of both tiled cusp and loop structures.
In Sec.~\ref{NLZ2}, we show NLZTs in the presence of only one loop structure. 
Finally, the conclusion follows in Sec.~\ref{conclusion}.

%%%%%%%%%%%%%%%%%%%%%%%%%%%%%%%%%%%%%%%%%%%%%%%%%%%%%%%%%%%%%%%%%%%%%%%%%%%%%%%%%%%%%%%%%%%
\begin{figure*}
\centering
\includegraphics[width=1\linewidth]{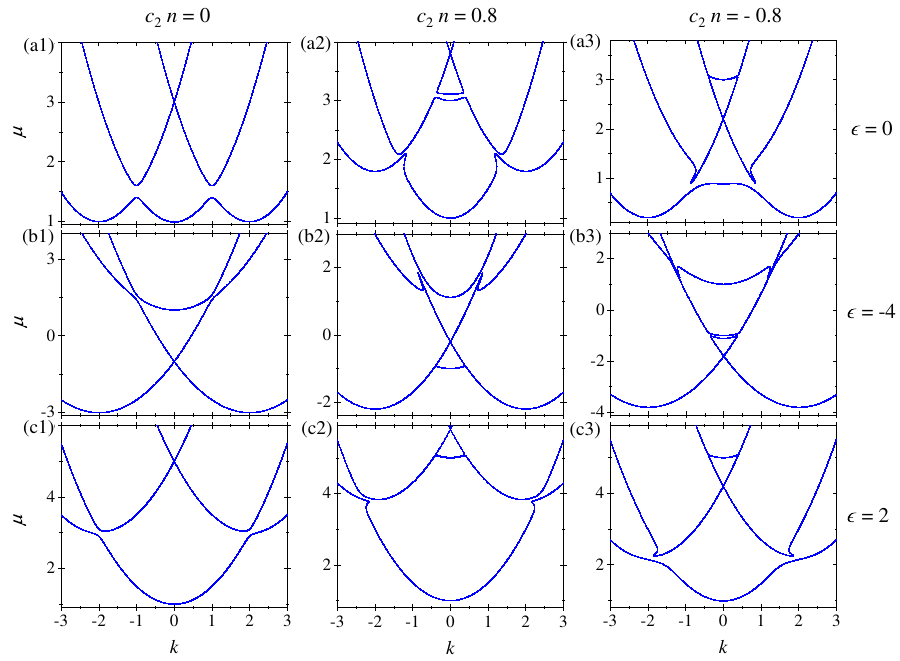}
\caption{Nonlinear dispersion relation of a spin-orbit-coupled spinor BEC. The Rabi frequency is $\Omega=0.1$, and the constant energy shift is $c_0n=1$. The quadratic Zeeman shift is set as $\epsilon=0$ in (a1)-(a3), $\epsilon=-4$ in (b1)-(b3), and $\epsilon=2$ in (c1)-(c3).
The first, second and third columns correspond to $c_2n=0$, $0.8$, and $-0.8$, respectively. In the single-particle dispersions in (a1), (b1) and (c1), there are three SOC avoided crossings. The quadratic Zeeman shift can tune the locations of the crossings in both $k$ and $\mu$ directions. The nonlinearity $c_2n\ne 0$ generates a loop in the middle avoided crossing and tilted cusps in the other two crossings. The bifurcations of the loop and cusps are determined by the sign of $c_2$.
}
\label{Fig1}
\end{figure*}
%%%%%%%%%%%%%%%%%%%%%%%%%%%%%%%%%%%%%%%%%%%%%%%%%%%%%%%%%%%%%%%%%%%%%%%%%%%%%%%%%%%%%%%%%%%

\section{Loop and tilted cusp featured nonlinear dispersion}
\label{dispersion}

The spin-1 spin-orbit coupling has been implemented in a spinor BEC by three Raman lasers~\cite{Campbell}. We consider this experimentally realizable system described by the following Hamiltonian, 
\begin{align}
H=\int d{x} \Psi^\dagger \left( H_{\text{SOC}}+
\frac{c_0}{2}\Psi^\dagger\Psi+\frac{c_2}{2}\Psi^\dagger\bm{F}\Psi\cdot\bm{F}
\right)\Psi,
\end{align} 
where $\Psi=(\Psi_{1},\Psi_{2},\Psi_{3})^T$ is the spin-1 spinor wave function with $\Psi_i$ being the wave functions of the $i$-th hyperfine state $\ket{i}$ ($i=1,2,3$), which satisfy the normalization condition $\int dx \Psi^\dagger\Psi =N$ with $N$ being the total atom number. $\bm{F}=(F_x,F_y,F_z)$ are the spin-1 Pauli matrices. The spin-1 interactions include the density-density part with the coefficient $c_0$ and  the spin-spin part with the coefficient $c_2$~\cite{Stamper-Kurn}. The single-particle SOC Hamiltonian is~\cite{Campbell,Sun,Yu,Martone,Lan}
\begin{equation}
H_\text{SOC}=\frac{1}{2}\left(-i\frac{\partial}{\partial x}+2F_z \right)^2 +\sqrt{2}\Omega F_x + \epsilon F^2_z.
\end{equation}
Here, $\Omega$ is the Rabi frequency, which originates from two-photon transitions induced by the Raman lasers, $\epsilon$ is the quadratic Zeeman shift, which can be changed by tuning the frequencies of the Raman lasers.  The Hamiltonian $H$ is dimensionless. The units of momentum, length, and energy are $\hbar k_\mathrm{Ram}$, $1/k_\mathrm{Ram}$, and $\hbar^2k^2_\mathrm{Ram}/m$, respectively, where $m$ is the atom mass and $k_\mathrm{Ram}=2\pi/\lambda_\mathrm{Ram}$ is the wave number of the Raman lasers with $\lambda_\mathrm{Ram}$ being the wavelength. Due to these units, the spin-orbit coupling strength becomes a fixed number $2$ in the front of $F_z$ in $H_\text{SOC}$.  By applying  
$i\partial\Psi(x)/\partial t=\delta H/\delta \Psi^\dagger(x)$, we obtain the Gross-Pitaevskii (GP) equation which describes dynamics of the SOC spinor BEC,
\begin{equation}
\label{GPE}
i\frac{\partial\Psi}{\partial t}=\left( H_\text{SOC}+H_\text{int}\left[\Psi\right]\right) \Psi,
\end{equation}
with the interacting part being
\begin{equation}
H_\text{int}\left[\Psi\right]=c_0\Psi^\dagger\Psi+c_2\Psi^\dagger\bm{F}\Psi\cdot\bm{F}.
\end{equation}

NLZTs are related to nonlinear dispersion relations. 
To obtain the nonlinear dispersion, we consider a spatially homogeneous BEC and  assume Eq.~\eqref{GPE} having plane-wave solutions,
\begin{equation}
\Psi =\sqrt{n} \exp(ikx - i\mu t)\Phi,
\end{equation}
where $n$ is the mean density, $k$ is the quasimomentum, and $\mu$ denotes the chemical potential. 
Then the spinor $\Phi = \left(\Phi_1 ,\Phi_2,\Phi_3\right)^T$, which is independent of $t$ and $x$, satisfies the stationary GP equation,
\begin{equation}
\label{NLE}
\mu \Phi = \left\lbrace  H_\text{SOC}\left(k\right) + H_\text{int}\left[\Phi\right]\right\rbrace  \Phi,
\end{equation}
and the normalization condition becomes $|\Phi_1|^2+|\Phi_2|^2+|\Phi_3|^2=1$. In Eq.~\eqref{NLE}, the interacting Hamiltonian is
\begin{equation}
H_\text{int}[\Phi]=c_0n+c_2n\Phi^\dagger\bm{F}\Phi\cdot\bm{F}.
\end{equation}
Therefore, the nonlinear coefficients are dressed with the mean density $n$, and the density-density interaction becomes a constant energy shift due to the feature of the plane-wave solutions.
The SOC Hamiltonian becomes
\begin{equation}
H_\text{SOC}(k)=\frac{(k+2F_z)^2}{2}+\sqrt{2}\Omega F_x + \epsilon F^2_z.
\end{equation}
The nonlinear dispersion relation $\mu(k)$, which is the dependence of the chemical potential on the quasimomentum, can be obtained by numerically solving Eq.~\eqref{NLE}. The density-density interactions $c_0n$ only contribute to an overall shift in the dispersion, and therefore do not have essentially physical consequences. The nonlinear effect is solely determined by the spin-spin interactions $c_2n$ which can be tuned by varying the mean density $n$ and by choosing different kinds of atoms to change the sign of $c_2$.

We first study the dispersions with tunable $c_2n$ in the absence of the quadratic Zeeman effect ($\epsilon=0$), which are demonstrated in Figs.~\ref{Fig1}(a1)-\ref{Fig1}(a3). 

(i) Figure~\ref{Fig1}(a1) shows the single-particle dispersion ($c_2n=0$). The dispersion has three bands, which are symmetric with respect to $k=0$ due to the symmetry $\mathcal{P}F_x$ with the parity operator $\mathcal{P}k \mathcal{P}^{-1}=-k$. There are three SOC avoided crossings; the left (right) one located around $k=-1$ ($k=1$) originates from the two-photon transition that couples $\Phi_1$ and $\Phi_2$ ($\Phi_2$ and $\Phi_3$), and therefore the gap sizes of these two crossings are proportional to the two-photon coupling strength $\Omega$;
the middle crossing around $k=0$ originates from the four-photon transition that couples $\Phi_1$ and $\Phi_3$, so its gap size is proportional to $\Omega^2$ [which is invisible in Fig.~\ref{Fig1}(a1)]. The left and right ones are symmetric with respect to $k=0$ and have a lower energy than the middle one. 

(ii) For an antiferromagnetic spin-spin interaction $c_2n=0.8$, the nonlinear dispersion is depicted in Fig.~\ref{Fig1}(a2). The nonlinear dispersion is still symmetric with respect to $k=0$. The outstanding feature is the appearance of loop-like structures around the SOC avoided crossings. Tilted cusp structures bifurcating from the lowest band appear in both the left and right avoided crossings, and a loop adhering to the highest band appear in the middle avoided crossing. We emphasize that the two cusps and single loop are purely nonlinear effects which will lead to intriguing NLZTs.

(iii) The nonlinear dispersion for a typical ferromagnetic spin-spin interaction $c_2n=-0.8$ is depicted in Fig.~\ref{Fig1}(a3). In comparison with Fig.~\ref{Fig1}(a2), both the tiled cusps and the loop bifurcate from the middle band.

\begin{figure}[t]
\centering
\includegraphics[width=1\linewidth]{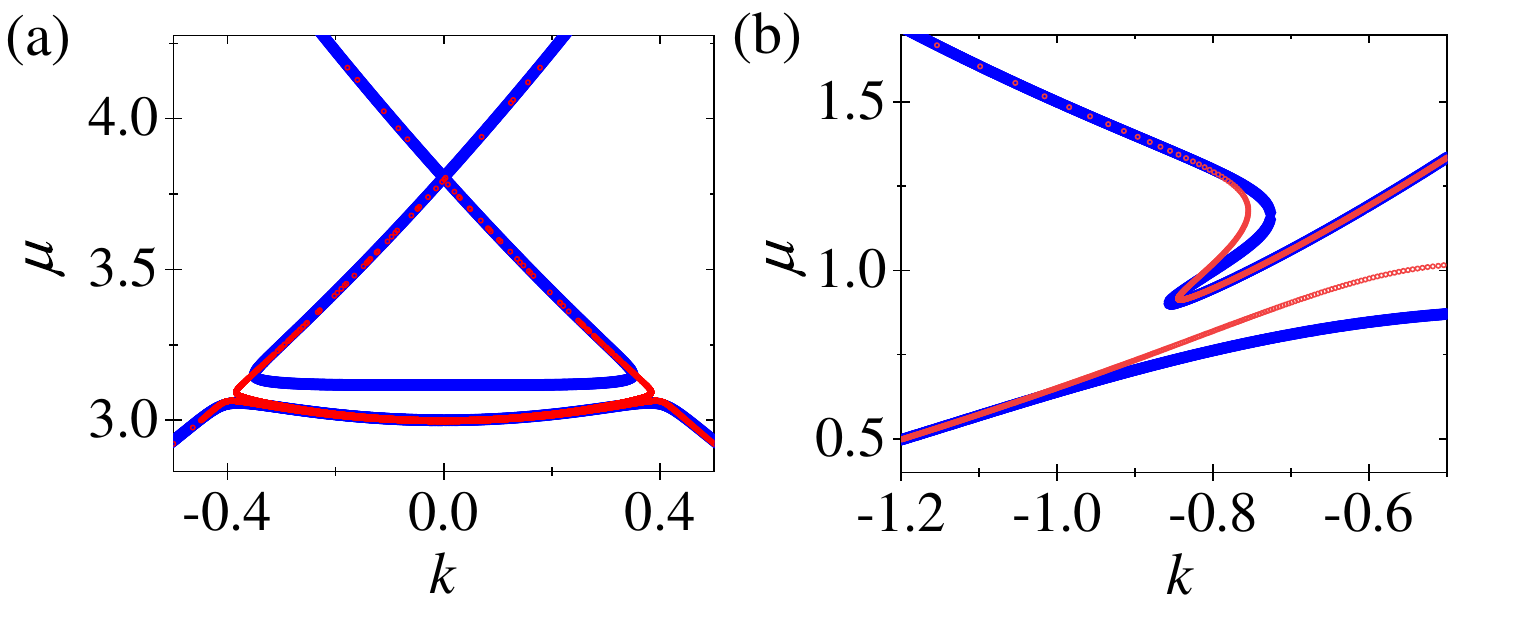}
\caption{Nonlinear dispersion of the effective two-level model [shown by the red-dotted line]. The blue lines are resultant dispersion from the spinor BEC in Eq.~\eqref{NLE}.  The parameters for (a) and (b) are same as those in Figs.~\ref{Fig1}(a2) and \ref{Fig1}(a3), respectively.}
\label{Fig2}
\end{figure}

The physical origination of these nonlinear structures can explain which band they bifurcate from and why they show a cusp or loop geometry. The middle crossing around $k=0$ originates from the coupling between $\ket{1}$ and $\ket{3}$.
By  assuming $\partial\Phi_2/\partial t=0$ and $\partial^2\Phi_2/\partial x^2=0$  in Eq.~\eqref{GPE}, we obtain $\Phi_2 \approx - \Omega\left(\Phi_1 + \Phi_3\right) /[ c_0n+c_2n(|\Phi_1|^2+|\Phi_3|^2+ 2\Phi_1\Phi_3 ) ]$, which is small quantity so that $\ket{2}$ can be adiabatically eliminated. 
In this way, the coupling between $\ket{1}$ and $\ket{3}$ can be shown explicitly.
By substituting $\Phi_2$ into  Eq.~\eqref{NLE} and neglecting the higher order terms of $\Phi_2$, we obtain the effective two-level model,
\begin{equation}
\mu_{\text{ntl1}}\Phi^\prime=H_{\text{ntl1}}\Phi^\prime.
\end{equation}
Here, $\Phi^\prime=(\Phi_1,\Phi_3)^T$ satisfies the normalization condition $|\Phi_1|^2+|\Phi_3|^2=1$, and $\mu =\mu_{\text{ntl1}}+k^2/2+2+\Omega^\prime+c_0n+\epsilon$ with
\begin{align}
\Omega^\prime &=  - \frac{\Omega^2}{c_{0}n + c_{2}n \left(1+2\Phi_{1}\Phi_{3}\right)},\notag  \\
 &  \approx   - \frac{\Omega^2}{c_{0}n + 2c_{2}n }.\notag 
 \end{align}
 We have assumed that $2\Phi_{1}\Phi_{3} \approx |\Phi_1|^2+|\Phi_3|^2=1$ to achieve a  wave-function independent $\Omega'$. 
The effective two-level Hamiltonian can be written as
\begin{equation}
\label{EffectiveOne}
H_{\text{ntl1}}=\Omega^\prime\sigma_x+2k\sigma_z - c_2n\left(\left|\Phi_{3}\right|^2 - \left|\Phi_{1}\right|^2\right)\sigma_z,
\end{equation}
where $\sigma_{x,z}$ are the spin-1/2 Pauli matrices.  The effective two-level model is similar to the standard two-level nonlinear system from which the existence and features of loop structure are analytically identified~\cite{LiuJ}. The terms $\Omega^\prime\sigma_x+2k\sigma_z$ define an avoided crossing centered at $k=0$ with the gap size being $2|\Omega'|$. If the nonlinearity dominates over the gap, i.e., $ |c_2|n> |\Omega'| $, a loop structure appears. 
When the effective interaction is attractive ($c_2>0$), the loop bifurcates from the upper band of the avoided crossing and grows downwards into the gap~\cite{LiuJ,Zhang2019}. When the effective interaction is repulsive ($c_2<0$), the loop adheres to the lower band of the avoided crossing and grows upwards into the gap.
According to Ref.~\cite{LiuJ}, the loop extends in the region $2|k|< |(c_2n)^{2/3} - \Omega^{\prime2/3}|^{3/2} $. 
In Fig.~\ref{Fig2}(a), we show the nonlinear dispersion of the effective two-level model for an attractive interaction $c_2n=0.8$. Generally, the dispersion of the effective model can match that from full calculation of the spinor BEC. However, a slight mismatch still exists due to the adiabatic elimination of $\Phi_2$.

%%%%%%%%%%%%%%%%%%%%%%%%%%%%%%%%%%%%%%%%%%%%%%%%%%%%%%%%%%%%%%%%%%%%%%%%%%%%%%%%%%%%%%%%%%%
%%%%%%%%%%%%%%%%%%%%%%%%%%%%%%%%%%%%%%%%%%%%%%%%%%%%%%%%%%%%%%%%%%%%%%%%%%%%%%%%%%%%%%%%%%%

The left avoided crossing around $k=-1$ shown in Figs.~\ref{Fig1}(a1)-\ref{Fig1}(a3) originates from the coupling between $\ket{1}$ and $\ket{2}$ and meanwhile has a negligible occupation of $\ket{3}$. By setting $\Phi_3=0$, Eq.~(\ref{NLE}) becomes the following effective two-level model,
\begin{equation}
\label{EffectiveTwo}
\mu_{\text{ntl2}}\Phi^{\prime\prime}=  H_\text{ntl2} \Phi^{\prime\prime},
\end{equation}
with $\Phi^{\prime\prime}=(\Phi_1,\Phi_2)^T$ satisfying $|\Phi_1|^2  +|\Phi_2|^2=1 $ and $\mu =\mu_{\mathrm{ntl2}}+k^2/2+k +1+\epsilon/2+c_0n+c_{2}n/2$. The corresponding effective  Hamiltonian is
\begin{equation}
\label{eq11}
H_{\text{ntl2}} =\Omega\sigma_x + \left(k +1 + \frac{\epsilon+c_2n}{2}\right)\sigma_z + 
\frac{c_{2}n}{2}\left|\Phi_{1}\right|^2  \left(I-\sigma_z\right),
\end{equation}
with $I$ being the $2\times2$ identical matrix.
In this case, the linear part of $H_{\text{ntl2}}$ represents an avoided crossing centered at $k=-1-(\epsilon+c_2n)/2$ with the gap of size $2\Omega$. The nonlinear term $c_2n|\Phi_1|^2(I- \sigma_z)/2$ is not symmetric as that in Eq.~(\ref{EffectiveOne}). 
Due to the asymmetric nonlinearity, the nonlinear structure is no longer a loop. 
The non-commutation of $\Omega \sigma_x$ and $-c_2n/2|\Phi_1|^2\sigma_z$ gives rise to the nonlinear cusp when $|c_2|n/2>\Omega$. In comparison with the standard nonlinear two-level model~\cite{LiuJ}, we know that the effective interaction behaves as repulsive when $c_2>0$ and it is effectively attractive when $c_2<0$. In the presence of the repulsive (attractive) interaction, the cusp bifurcates from the lower (higher) band of the avoided crossing. Furthermore, the cusp structure is tilted due to the terms $k^2/2+k$ in the chemical potential.
In Fig.~\ref{Fig2}(b), we plot the nonlinear dispersion $\mu(k)$ calculated from the effective two-level mode in Eq.~(\ref{EffectiveTwo}) for an effectively attractive interaction $c_2n=-0.8$. 
It shows that the effective model can describe the tiled cusp very well.

%The right avoided crossing around $k=1$ shown in Figs.~\ref{Fig1}(a1)-(a3) originates from the coupling between $\Phi_2$ and $\Phi_3$. Similar as the left avoided crossing, one can study its features by building up a similar effective two-level model by assuming $\Phi_1=0$. 

The experimentally tunable quadratic Zeeman shift $\epsilon$ is of essential importance to modify the nonlinear dispersion. Figures~\ref{Fig1}(b1)-\ref{Fig1}(b3) demonstrate the dispersions for different nonlinearities with a negative quadratic Zeeman shift $\epsilon=-4$.  In the single-particle dispersion shown in Fig.~\ref{Fig1}(b1), the negative $\epsilon$ pulls down the components $\ket{1}$ and $\ket{3}$.
As a result, the middle avoided crossing at $k=0$ [which is still invisible due to the extremely small gap] declines and the right and left ones at $k=\pm 1$ are lifted, meanwhile the gap size of the left and right ones is obviously smaller than that of the $\epsilon=0$ case [see Fig.~\ref{Fig1}(a1)]. Finally, the middle one is located in the lower part compared with the right and left ones. The left (right) one is due to the coupling between $\ket{2}$ and $\ket{3}$ ($\ket{2}$ and $\ket{1}$). For the antiferromagnetic spin-spin interaction $c_2n=0.8$ demonstrated in Fig.~\ref{Fig1}(b2), tiled cusps appear in the right and left avoided crossings and a loop exists in the middle avoided crossing. Both the cusps and the loop bifurcate from the middle band. While, for the ferromagnetic spin-spin interaction $c_2n=-0.8$ [see Fig.~\ref{Fig1}(b3)], the tilted cusps bifurcate from the highest band and the loop bifurcates from the lowest band.

Figures~\ref{Fig1}(c1)-\ref{Fig1}(c3) describe the dispersions for different nonlinearities with a positive quadratic Zeeman shift $\epsilon=2$. In the single-particle dispersion shown in Fig.~\ref{Fig1}(c1), the positive $\epsilon$ pushes up $\Phi_1$ and $\Phi_3$. The left (right) avoided crossing still originates from the coupling between $\ket{1}$ and $\ket{2}$ ($\ket{2}$ and $\ket{3}$), but its location moves towards a larger $|k|$ around 2. The middle one is still at $k=0$. The nonlinear dispersions in Fig.~\ref{Fig1}(c2) [$c_2n=0.8$] and Fig.~\ref{Fig1}(c3) [$c_2n=-0.8$] show the appearance of tilted cusps and a single loop.

We can make following conclusions from Fig.~\ref{Fig1}. In a SOC spinor system, there are three SOC avoided crossings originating from the couplings between the two of three components. The coupling of $\ket{1}$ and $\ket{3}$ stemming from the four-photon transition induces the middle avoided crossing which is always located at $k=0$ with a higher order smaller gap size. The couplings of $\ket{1}$ and $\ket{2}$ (or $\ket{2}$ and $\ket{3}$) induced by the two-photon transitions give rise to the other two crossings located at a finite $|k|$. They are always symmetric with respect to $k=0$. The quadratic Zeeman shift $\epsilon$ can change the locations of the other two crossings in quasimomentum $k$ space and tune the relative chemical potential between the three crossings. The nonlinearity $c_2n\ne 0$ always generates a loop in the middle avoided crossing and tilted cusps in the other two. The loop bifurcates from the upper (lower) band of the middle crossing for $c_2n>0$ ($c_2n<0$). The tilted cusps bifurcate from the lower (upper) band of the other two avoided crossings for $c_2n>0$ ($c_2n<0$).

%%%%%%%%%%%%%%%%%%%%%%%%%%%%%%%%%%%%%%%%%%%%%%%%%%%%%%%%%%%%%%%%%%%%%%%%%%%%%%%%%%%%%%%%%%%
\begin{figure*}
\centering
\includegraphics[width=1\linewidth]{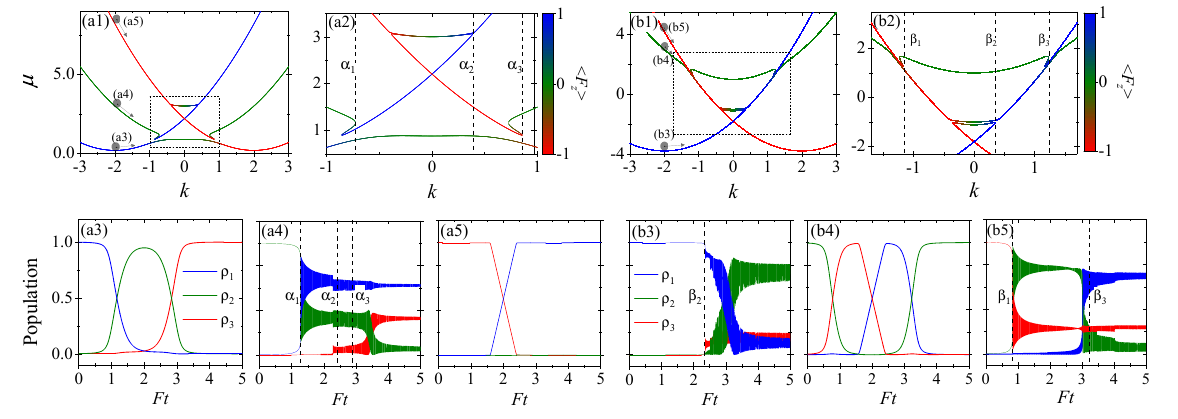}
\caption{Nonlinear dynamics induced by the weak acceleration force $F=0.0001$.  (a1) Nonlinear dispersion relation shown in Fig.~\ref{Fig1}(a3) which incorporates the color scale to show the tensor magnetization $\langle F_z \rangle=|\Phi_1|^2-|\Phi_3|^2$ of the corresponding eigenstates. Parameters are  $\epsilon=0$, $\Omega=0.1$, $c_0n=1$, and  $c_2n=-0.8$.  (a2) The zoom-in of the selected box region in (a1). 
(a3)--(a5) The time evolution of populations $\rho_i=|\Phi_i|^2$ for the different initial states which are labeled by the dots ``(a3)," ``(a4)," and ``(a5)" in Fig.~\ref{Fig1}(a1), respectively.
(b1)-(b5) show the same quantities as (a1)-(a5), but the nonlinear dispersion is modified by the quadratic Zeeman shift $\epsilon = -4$, and other parameters are $\Omega=0.1$, $c_0n=1$, and  $c_2n=-0.8$. }
\label{Fig3}
\end{figure*}
%%%%%%%%%%%%%%%%%%%%%%%%%%%%%%%%%%%%%%%%%%%%%%%%%%%%%%%%%%%%%%%%%%%%%%%%%%%%%%%%%%%%%%%%%%%

\section{NLZT in the presence of both the tilted cusps and the loop}
\label{NLZ}

The fundamental consequence of these nonlinear structures of tilted cusps and loops is the NLZT,
which can be observed by applying an acceleration force $F$ to move the BEC along the nonlinear dispersion~\cite{Zhang2019}. 
In experiments, the force may come from the gravitational force or the optical trap induced force in momentum space, the magnitude of which can be tuned by adjusting the trap frequency~\cite{Olson}. The force changes the quasimomentum of the BEC as 
\begin{equation}
\label{momentum}
  k(t)=k_\text{ini}+Ft,  
\end{equation}
with $k_\text{ini}$ being the initial quasimomentum. We stimulate nonlinear dynamics induced by a weak acceleration force $F=0.0001$ by numerically integrating Eq.~(\ref{GPE}) in momentum space with help from Eq.~(\ref{momentum}). The weak force is chosen in order to avoid the excitation induced by the normal Landau-Zener transitions.  

We first study motions along the nonlinear dispersion shown in Fig.~\ref{Fig1}(a3).  This dispersion is shown again in Fig.~\ref{Fig3}(a1) which incorporates the tensor magnetization $\langle F_z \rangle=|\Phi_1|^2-|\Phi_3|^2$ by the color. 
The blue, green and red colors correspond to eigenstates dominated by $\Phi_1$, $\Phi_2$, and $\Phi_3$ respectively.  
The nonlinear dynamics induced by the weak acceleration force strongly depend on the initial states which can be well prepared in experiments~\cite{Olson}.

(i) The initial state is chosen as the ground state at $k=-2$ which is labeled by the dot ``(a3)" in Fig.~\ref{Fig3}(a1). The weak acceleration force pushes atoms to move adiabatically along the lowest band, 
in the presence of a finite gap between the lowest two bands [see the zoom-in dispersion shown by Fig.~\ref{Fig3}(a2)]. The spin-momentum locking of the spin-orbit coupling is represented by the continuous color change in the lowest band. The time evolution of populations $\rho_i=|\Phi_i|^2$ are demonstrated in Fig.~\ref{Fig3}(a3). The initial ground state is mainly occupied by $\rho_1$. When atoms go through the left avoided crossing, $\rho_1$ is smoothly swapped with $\rho_2$, and around the right avoided crossing $\rho_2$ is smoothly swapped with $\rho_3$. These smooth exchanges of populations are the signature of the adiabatic motion along the lowest band.

(ii) The initial state is arranged at $k=-2$ in the middle band [as shown by the dot labeled as ``(a4)" in Fig.~\ref{Fig3}(a1)]. The dominant occupation in the initial state is $\rho_2$. The force moves the initial state towards the tilted cusp in the left avoided crossing. When atoms reach the edge of the cusp [represented by the vertical dashed line ``$\alpha_1$" in Fig.~\ref{Fig3}(a2)], there is no further trajectory for them to go forward. They must jump to the neighboring lowest two bands, leading to NLZT. The population $\rho_2$ suddenly decreases and $\rho_1$ dramatically increases at the time corresponding to $\alpha_1$ [see Fig.~\ref{Fig3}(a4)], and after these sudden changes both $\rho_2$ and $\rho_1$ oscillate fast. These sudden changes and fast oscillations are the distinguished features of NLZT. After the NLZT at the edge of the cusp, atoms cannot adiabatically follow a certain band and the emergent fast oscillations are due to the beating between the lowest two bands. Further acceleration pushes atoms to move along the bands represented by green and blue colors in Fig.~\ref{Fig3}(a2). The time evolutions of $\rho_2$ and $\rho_1$ have no obvious changes even when atoms reach the edge of the loop [the vertical line ``$\alpha_2$"] and the edge of the right cusp [the vertical line ``$\alpha_3$"]. There is no occupation of the red-colored band shown in Fig.~\ref{Fig3}(a2). After the time  corresponding to ``$\alpha_3$", there is a swapping between $\rho_2$ and $\rho_3$ [see Fig.~\ref{Fig3}(a4)], which indicates that some atoms move along the lowest band and pass the right avoided crossing.

(iii) The initial state is arranged at $k=-2$ in the highest band [as shown by the dot labeled as ``(a5)" in Fig.~\ref{Fig3}(a1)]. The dominant occupation in the initial state is $\rho_3$. The force pushes the initial state towards the middle avoided crossing. The loop bifurcates from the middle band and there is a very small gap between the loop and the highest band [which is invisible in Fig.~\ref{Fig3}(a2)]. Finally, atoms adiabatically move along the highest band. Since the population of $\rho_2$ is the negligible in the highest band, the time evolution of $\rho_2$ illustrated in Fig.~\ref{Fig3}(a5) always vanishes. 
Around the edges of middle avoided crossing at $k\approx 0.4$, the highest band has dramatic warps due to the existence of the loop, which makes the swapping between $\rho_3$ and $\rho_2$ not smooth [see  Fig.~\ref{Fig3}(a5)]. 

% Around the middle avoided crossing at $k=0$, the third band has dramatic warps due to the existence of the loop, which makes the swapping between $\rho_3$ and $\rho_2$ not smooth [see  Fig.~\ref{Fig3}(a5)]. 

Therefore, in order to stimulate NLZT, the initial state should be prepared in the same band from which the nonlinear structures bifurcate.  In the following, we study NLZT in the dispersion modified by the quadratic Zeeman effect, as shown in Fig.~\ref{Fig1}(b3). This dispersion is demonstrated in Fig.~\ref{Fig3}(b1) again with the color scale to guide the dominant population in eigenstates. The cusp and loop structures are zoomed-in in Fig.~\ref{Fig3}(b2). 

(i) Figure~\ref{Fig3}(b3) demonstrates the time evolution of populations for the initial state in the lowest band [labeled by the dot ``(b3)" in Fig.~\ref{Fig3}(b1)]. The force pushes atoms to the edge of the loop in the middle avoided crossing [represented by the vertical dashed line ``$\beta_2$"] and further push leads to NLZT. 
Around the NLZT, $\rho_1$ and $\rho_2$ change dramatically [see the vertical dashed line  ``$\beta_2$" in Fig.~\ref{Fig3}(b3)]. The dramatic changes and following fast oscillations in $\rho_1$ and $\rho_2$ signify that the NLZT results in the nonadiabatic superposition of the blue- and red-colored bands shown in Fig.~\ref{Fig3}(b2). In the further evolution, there is a swapping between $\rho_1$ and $\rho_2$ [see Fig.~\ref{Fig3}(b3)], since the blue-colored band crosses over to the green-colored band around the right avoided crossing at $k=\beta_3$ in Fig.~\ref{Fig3}(b2). 

(ii)  Figure~\ref{Fig3}(b4) illustrates the time evolution of populations for the initial state in the middle band [labeled by the dot ``(b4)" in Fig.~\ref{Fig3}(b1)]. The results indicate that the smooth swapping between $\rho_2$ and $\rho_3$ around $Ft=0.8$, $\rho_3$ and $\rho_1$ around $Ft=2$, and $\rho_1$ and $\rho_2$ around $Ft=3.4$ are due to the adiabatic motions along the band around the left, middle, and right avoided crossings. There is no NLZT happening in this case. 

(iii)  Figure~\ref{Fig3}(b5) shows the time evolution of populations for the initial state in the highest band [labeled by the dot ``(b5)" in Fig.~\ref{Fig3}(b1)]. There are two NLZTs around $Ft=\beta_1$ and  $Ft=\beta_3$ which correspond to the edges of the cusps in the left and right avoided crossings [see vertical lines in  Fig.~\ref{Fig3}(b2)]. After the first NLZT, atoms occupy the green and red colored bands shown in Fig.~\ref{Fig3}(b2). The population of the red-colored band adiabatically evolves along this band so that there is no physical change when $Ft>\beta_1$. However, the population of the green-colored band experiences the second NLZT when atoms reach the edge $\beta_3$ of the cusp in the right avoided crossing.

\section{NLZT in the presence of the loop}
\label{NLZ2}

%%%%%%%%%%%%%%%%%%%%%%%%%%%%%%%%%%%%%%%%%%%%%%%%%%%%%%%%%%%%%%%%%%%%%%%%%%%%%%%%%%%%%%%%%%%
\begin{figure}
\centering
\includegraphics[width=0.8\linewidth]{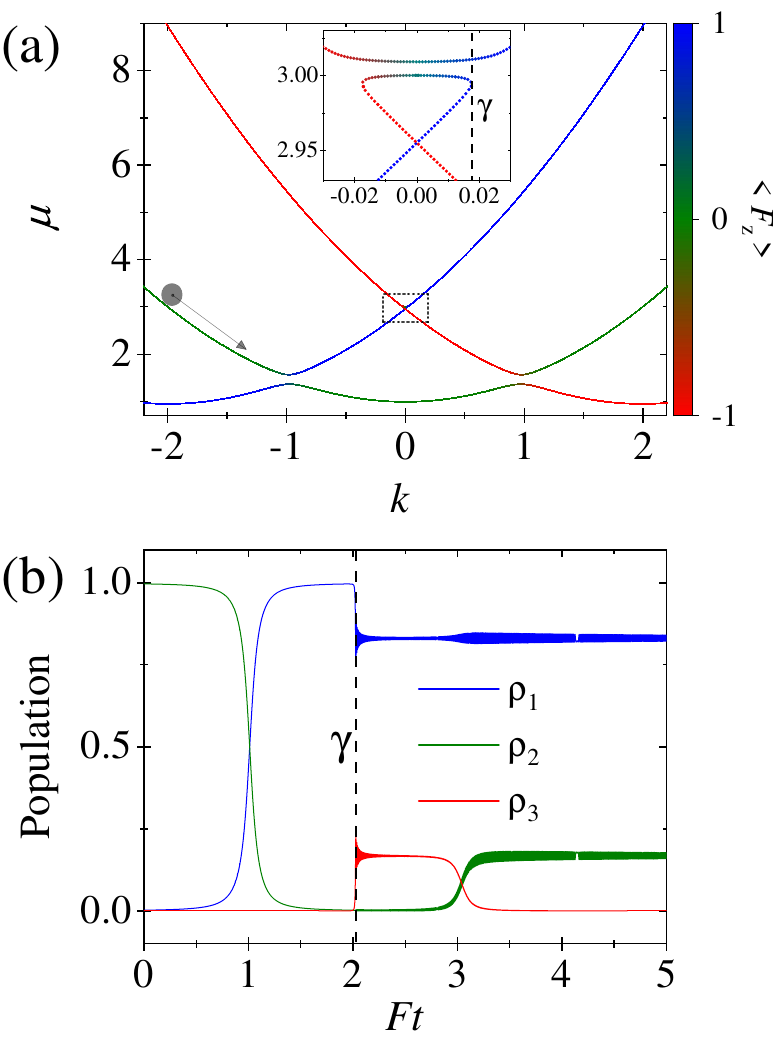}
\caption{NLZT in the presence of only one loop in the middle avoided crossing, with $c_2n=-0.05$.  (a) Nonlinear dispersion with the color scale to show the tensor magnetization of corresponding eigenstates.  The inset is the zoom-in of the looped avoided crossing. 
(b) The dynamics of populations with the initial state shown by the dot in 
(a), which is triggered by the weak acceleration force $F=0.0001$. 
Other parameters are same as in Fig.~\ref{Fig3}(a1).}
\label{Fig4}
\end{figure}
%%%%%%%%%%%%%%%%%%%%%%%%%%%%%%%%%%%%%%%%%%%%%%%%%%%%%%%%%%%%%%%%%%%%%%%%%%%%%%%%%%%%%%%%%%%

We have shown the possible NLZTs induced by the nonlinear cusp and loop structures. 
The appearance of these structures requires the nonlinearity $c_2n$ to dominate gap sizes of avoided crossings. In the  SOC spin-1 system, the four-photon transition generates an SOC avoided crossing with the gap much smaller than the two-photon transition induced ones. Therefore, there exists the nonlinear regime, in which a loop appears in the four-photon-induced avoided crossing, and there is no cusp in the two-photon-induced ones. 
Figure~\ref{Fig4}(a) demonstrates such a nonlinear dispersion with $c_2n=-0.05$, in which  the loop structure bifurcating from the middle band is zoomed-in. In order to observe NLZT, the initial state should be prepared in the same band, which is labeled by the dot in Fig.~\ref{Fig4}(a).  The weak acceleration force pushes the initial state to move along the middle band. When atoms pass the left avoided crossing, the populations $\rho_2$ and $\rho_1$ are swapped [see Fig.~\ref{Fig4}(b)]. The further acceleration pushes atoms to the edge of the loop [represented by the vertical dashed line ``$\gamma$" in the inset of Fig.~\ref{Fig4}(b)]. Then the NLZT happens, which is reflected by the sudden changes of populations $\rho_1$ and $\rho_3$, and by the following fast oscillations in Fig.~\ref{Fig4}(b). 
After the transition at $Ft=\gamma$, the populations $\rho_1$ and $\rho_3$ are occupied, illustrating that atoms stay in the bands represented by red and blue colors in Fig.~\ref{Fig4}(a). The atoms in the red-colored band further have a swapping between $\rho_3$ and $\rho_2$ when they are accelerated around the right avoided crossing.

The NLZT shown in Fig.~\ref{Fig4} needs only a small nonlinearity $c_2n$, which is preferable for its experimental realization in $^{87}$Rb spinor BECs.

\section{Experimental feasibility}

We have shown rich NLZT dynamics in SOC spinor BECs in previous sections. Now we discuss its experimental feasibility in such a system. A SOC spinor BEC can be experimentally implemented in $^{87}$Rb atoms by multiple two-photon couplings induced via three Raman lasers~\cite{Campbell} or two Raman lasers~\cite{Olson}. Especially, in the three-Raman-laser experiment, the synthetic quadratic Zeeman shift is experimentally tunable~\cite{Campbell}. This experimentally realizable system has multiple energy avoided crossings in the dispersion relation.  The gap sizes of the avoided crossings relate to the Rabi frequency $\Omega$ which can be experimentally adjusted by changing the intensity of Raman lasers. 

The spinor BEC is characterized by the density-density interactions with the strength $c_0n$ and the spin-spin interactions with $c_2n$. In $^{87}$Rb BECs, $|c_2|$ is much smaller than $c_0$, so that most of physics are dominated by the density-density interactions. Very interestingly, the emergence of tilted cusp and loop structures around the avoided crossings is only relevant to $c_2n$, and $c_0n$ does not contribute to it. This fundamental property enables experiments to enhance $c_2n$ by changing atomic density $n$ which can be implemented by either tightening the transversal traps or increasing atom number. The initial states in different bands can be precisely prepared by controlling populations on hyperfine states~\cite{Olson,Olson2017}. Then, they are accelerated to move along the nonlinear dispersion by the acceleration force which may come from the gravitational force or the longitudinal optical trap~\cite{Olson}. Especially, the magnitude of the force induced by the longitudinal optical trap is tunable by adjusting the trapping frequency~\cite{Olson}. With the acceleration force, the momentum of BECs changes linearly as function of time. The nonlinear dispersion exhibits spin-momentum locking originating from the SOC. 
Linearly changing momentum alters the spin polarization simultaneously. 
Through the time and spin resolved measurement of spin polarization, NLZT can be observed time-dependently.

\section{Conclusion}
\label{conclusion}

We have shown that a SOC spinor BEC is a tunable and versatile platform to study NLZT. The NLZTs can be induced by nonlinear cusp and loop structures appearing around SOC avoided crossings in the nonlinear dispersion.  The emergence of these nonlinear structures is only related to the spin-spin interactions and does not depend on the density-density interactions. This makes it possible to tune the nonlinearity by adjusting atomic density and to observe NLZT in a $^{87}$Rb spinor BEC with the spin-orbit coupling. 
Our study provides an interesting route to investigate nonlinear dynamics in spinor BECs.

\section*{Acknowledgment}

We acknowledge valuable discussions with Prof. Doerte Blume and Prof. Yong Xu. This work is supported by National Natural Science Foundation of China with Grants No. 12374247 and No. 11974235, and Shanghai Municipal Science and Technology Major Project (Grant No. 2019SHZDZX01-ZX04). 
L.H. also acknowledges support from Okinawa Institute of Science and Technology Graduate University.

\bibliography{spin1NLZ}

\end{document}